\documentclass[prb,floatfix,twocolumn,showpacs,amsfonts]{revtex4}
\usepackage{graphicx}
\usepackage{bm}
\begin{document}
\title{Theory of $d$-density wave  viewed from  a vertex model and its implications}
\author{Sudip Chakravarty}
\email[Electronic address:]{sudip@physics.ucla.edu}

\affiliation{Department of Physics and Astronomy, University of California Los
Angeles, Los Angeles, California 90095}
\date{\today}
\begin{abstract}
The thermal disordering  of the $d$-density wave, proposed to be the origin of the pseudogap state of high temperature
superconductors, is suggested to be the same as that of the statistical mechanical model known as the 6-vertex model.
The low temperature phase consists of a staggered order parameter of circulating currents, while the disordered high
temperature phase is a power-law phase with no order. A special feature of this transition is the complete lack of an
observable specific heat anomaly at the transition. There is  also a transition at  a even higher temperature at which
the  magnitude of the order parameter collapses. These  results are due  to classical thermal fluctuations and are 
entirely unrelated to a quantum critical point in the ground state. The quantum mechanical ground state can be explored
by incorporating processes that causes transitions between the vertices, allowing us to discuss quantum phase
transition in the ground state as well as the effect of quantum criticality at a finite temperature as distinct from
the power-law fluctuations in the classical regime. A generalization of the model on a triangular lattice that leads to
a 20-vertex model may shed light on the Wigner glass picture of the metal-insulator transition in two-dimensional electron
gas. The
 power-law ordered high temperature phase may be generic to a class of
constrained systems  and its relation to recent advances in the quantum dimer models is noted. 
\end{abstract}
\pacs{}
\maketitle

\section{Introduction}
A motivating factor for the present paper is the specific suggestion that a  new broken symmetry
can explain the  pseudogap
phase of  high temperature superconductors\cite{Chakravarty1,Chakravarty2}. The corresponding order is
a particle-hole condensate of ``angular momentum" 2, termed the $d$-density wave (DDW)\cite{DDW}. There
is some tell-tale evidence of this unusual order parameter\cite{Mook} involving circulating orbital
currents arranged in a staggered pattern, which is directly detected as a Bragg scattering signal in neutron measurements. 
The second motivating factor is to place this proposal in a wider context of many body theory, where a
strongly correlated electronic system can have unconventional broken symmetries in the ground state. In this respect, I shall
briefly touch upon the topic of two-dimensional electron gas in Si-MOSFET devices\cite{Abrahams}.

The building blocks of the low energy theory corresponding to DDW are bond currents whose arrangements define the
various order parameters that reflect particle-hole condensates\cite{Nayak}. The idea is clearly similar to  resonating
valence bonds (RVB)\cite{Anderson}, where the building blocks are valence bonds  that can be described in terms of
particle-particle condensates. These  may order in the ground state, or they may not, in which case one has a spin
liquid\cite{Sondhi}. In either case,  strong correlation effects are believed to play an important role. The
parallel goes further: while a tractable RVB Hamiltonian is the quantum dimer model\cite{Kivelson}, a  tractable model
for bond currents will be seen to be a vertex model known in statistical mechanics and its suitable quantum
generalization.  

The actual statistical mechanics of the DDW 
transition is richer than the mean field (Hartree-Fock) picture in which the ordered pattern is frozen,
until the magnitude of the circulating currents vanishes, which is the simplest possible description of
the broken symmetry phase. While this is reasonable deep in the ordered state, it does not allow
for  fluctuations of the order parameter.

A natural modification that I shall describe,  involving the 6-vertex model\cite{Lieb,Baxter},   leads to striking
consequences for  the pseudogap phase: (1) As the
$d$-density wave disorders with increasing temperature, the system exits to a  power-law
phase that is not due to  any underlying quantum critical points.  (2)
The pseudogap transition does not have any specific heat anomaly\cite{Kee}. (3) Because of the power-law
nature of the orbital current correlations of the disordered state, it is likely that the electron
spectral function exhibits a cut spectrum, unlike a Fermi liquid. (4)  As the temperature is raised, a
second  transition takes place at a higher temperature, where the local amplitude of the pseudogap
vanishes. 
It is unusual, but true, that the two-dimensional classical statistical mechanics of the pseudogap state
viewed as a DDW allows for a  power-law phase with infinite correlation length in the high
temperature regime, while the low temperature phase is ordered with a finite correlation length.

The outline of the paper is as follows. In Sec. II, I introduce and discuss the 6-vertex model for the pseudogap phase, and, in Sec.
III, I discuss its qunatum generalization. Section IV is a brief analysis of the relation with the quantum dimer model. In Sec. V,
I introduce  the 20-vertex model on a triangular lattice and discuss its relevance to the Wigner glass picture of the in 
metal-insulator transition.  And,
finally, the conclusions are summarized in Sec. VI. 

\section{The 6-vertex model for the pseudogap phase}
\begin{figure}[htb]
\centerline{\includegraphics[scale=0.35]{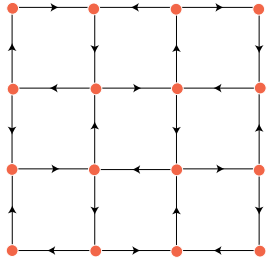}}
\caption{A segment of the circulating current pattern in the DDW ordered state.}
\label{fig:DDW}
\end{figure}

\begin{figure}[htb]
\centerline{\includegraphics[scale=0.25]{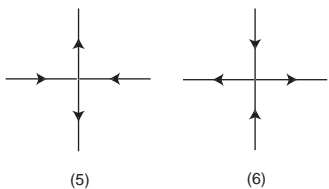}}
\caption{Vertices 5 and 6.}
\label{fig:vertex1}
\end{figure}

The ordered singlet DDW state consists of a staggered circulating  pattern of currents
flowing on the square planar CuO-lattice, as shown in Fig.~\ref{fig:DDW}.
The configuration  is  a result of juxtaposing two sets of vertices, shown in Fig.~\ref{fig:vertex1}, centered on the Cu
atoms in a current conserving manner.
There are only two possible choices, resulting in two distinct broken symmetry states,
which are related by time reversal and a lattice translation. 

More mathematically, the order parameter $\Phi_{\bf Q}$ is given by
\begin{equation}
\langle {\psi^{\sigma\dagger}}({\bf k}+{\bf Q},t)
{\psi_\rho}({\bf k},t)\rangle
= i{\frac{\Phi_{\bf Q}}{2}}\,(\cos{k_x}-\cos{k_y})\, {\delta^\sigma_\rho},
\end{equation}
where $\sigma$ and $\rho$ are the spin indices, and ${\mathbf Q}=(\pi,\pi)$.  The lattice
constant has been set to unity, and the operator $\psi_\rho$ is the electron destruction operator.
The order parameter is called a density wave because it is
a particle-hole condensate, even though what is actually modulated is current, not density. The order parameter $\Phi_{\bf Q}$ is 
a spin singlet; there is also a triplet version in which it is the spin current that is modulated.\cite{Nayak} The order parameter
is called a $d$-wave because of the internal form factor of the particle and  the hole, which is $(\cos{k_x}-\cos{k_y})$. Of course,
on a crystalline lattice, angular momentum is not a good quantum number. This is the closest we can come to the angular momentum 2
of a
$d_{x^2-y^2}$ wave function in free space. Note that for particle-hole
condensates there are no exchange requirements governing order parameter symmetries that enslave the
orbital wave function to the spin wave function, in contrast to superconductivity where the condensate is of particle-particle type.
Thus, one can have a DDW that is either a spin singlet, or a spin triplet.
 
In a mean field (Hartree-Fock) picture, the only way the state can
disorder  is by the thermal collapse of the magnitude of the order parameter, which  is a highly
restrictive mechanism   because it does not allow for fluctuations. The system may disorder long
before the magnitude of the order parameter collapses. 
To build fluctuation effects, we consider the
basic building blocks that are  bond currents between the nearest neighbor sites $\bf x$ and ${\bf
y}={\bf x}\pm \hat x$, or ${\bf y}={\bf x}\pm \hat{y}$, defined by
\begin{equation}
{\mathbf j}_{\bf x,y} = -i\epsilon \Phi_{\bf Q}\langle \psi^\dagger_{\bf x}\psi_{\bf y}-\text{h. c.} \rangle, 
\label{eq:current}
\end{equation}
where $\epsilon =\pm 1$ determines the direction of the current flow. The status of the bond currents
is identical to the local order parameter  in Ginzburg-Landau-Wilson formalism in which the
partition function is a sum over all complexions of the order parameter weighted by the effective coarse
grained action. To include quantum fluctuations, one must supplement the theory with a suitable dynamics, which I shall discuss in
the following section. 

I shall assume that there is  a regime of coupling constant and temperature
such that the bond current order is well developed, and the fluctuations of the magnitudes of the bond
currents can be neglected. If this is the case, there should be a second transition at which the magnitude of the
bond current itself vanishes. This would  imply a second pseudogap transition at a
higher temperature. The existence of a specific heat anomaly at this upper pseudogap transition is not entirely clear, as it will
be seen to be a transition from a power-law ordered phase. However, in any case, the temperature may be too large to extract it from
the large phonon or other backgrounds. I shall therefore ignore this transition altogether.

It is
easy to convince oneself that  low lying thermal or quantum fluctuations can be expected  to reverse
a set of bond currents (without changing their magnitudes) provided  no sources or sinks are generated,
that is, ${\mathbf \nabla}\cdot{\mathbf j}=0$. Given that there are two incoming and
two outgoing currents at a vertex of a square lattice, there are altogether $\frac{4!}{2!2!}=6$ possible vertices. Thus,
there are four  additional allowed vertices, beyond the two shown in
Fig.~\ref{fig:vertex1}. These are shown below in Fig.~\ref{fig:vertex2}. 
\begin{figure}[htb]
\centerline{\includegraphics[scale=0.25]{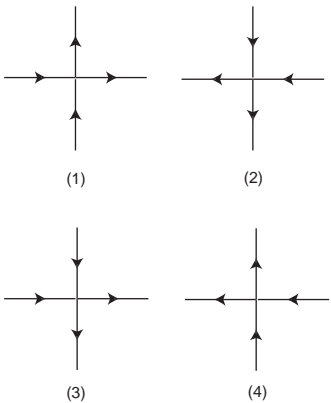}}
\caption{Vertices 1, 2, 3 and 4.}
\label{fig:vertex2}
\end{figure}
In general, each of the 6 local arrangements can have a distinct energy $\varepsilon_i$, but if we impose the
restriction that there are no net external currents, then
\begin{equation}
\varepsilon_1 = \varepsilon_2, \; \varepsilon_3 = \varepsilon_4, \; \varepsilon_5 = \varepsilon_6.
\end{equation}
The model is then unchanged by reversing all the arrows. For tetragonal symmetry, there are only two independent
energy scales
$\varepsilon_1$ and
$\varepsilon_5$ because $\varepsilon_1 =\varepsilon_3$, but for generality we shall assume that they are
distinct. 

From a  Hartree-Fock analysis,\cite{Nayak} it can be argued that $\varepsilon_1$ and $\varepsilon_5$ are close in energy. 
This is because the vertices $1-4$ correspond to ($p_x+p_y$)-density wave (PDW) states in the order parameter language,
and the energetics of both the singlet $p$-density wave and the DDW are controlled  by small pair-hopping matrix elements,
smaller than the scale of the antiferromagnetic exchange constant $J$.  A singlet $p_x$-density wave state has the ordering
\begin{equation}
\langle {\psi^{\sigma\dagger}}({\bf k}+{\bf Q'},t)
{\psi_\rho}({\bf k},t)\rangle
= \Phi'_{\bf Q'}\, \sin k_x\, {\delta^\sigma_\rho}.
\end{equation}
where ${\bf Q}'=(0,\pi)$. Note that in this case $\Phi'_{\bf Q'}$ is real, but because of the form factor $\sin k_x$, 
a Fourier transform to the real space brings out the pattern of currents shown in vertices  in Fig.~\ref{fig:vertex2}, when it
is superposed with the corresponding $p_y$-density wave.  The order parameter
$\Phi'_{\bf Q'}$ is the closest analog on a lattice of angular momentum 1,
$p_x$ wave function in free space. Note, as before,  that it is the current that is modulated, not the density.   

I shall first consider thermal fluctuations and take into account quantum fluctuations in the following section. A basic assumption
I shall make is that the energy of a given state is a simple sum of energies associated with the configuration at each vertex.
This is a reasonable assumption because any long-ranged interaction between the vertices are unlikely, as the thermal smearing will
cause any interactions mediated by the nodal quasiparticles to be exponentially decaying.  Following the conventional notation, I
shall define, 
\begin{equation}
a=\omega_1=\omega_2, \: b=\omega_3=\omega_4, \; c=\omega_5=\omega_6,
\end{equation}
where the Boltzmann factors are defined by $\omega_i=e^{-\varepsilon_i/T}$, where the Boltzmann constant $k_B$ is set
to unity. The partition function is
\begin{equation}
Z=\sum a^{n_1+n_2} b^{n_3+n_4} c^{n_5+n_6},
\end{equation}
where $n_k$ is the number of vertices of type $k$. The  sum is over all
arrangements that fit together continuously without generating sources and sinks. The partition function
is precisely the partition of the 6-vertex model for which many exact results are known\cite{Lieb,Baxter}.

\subsection{Phase diagram}
\begin{figure}[htb]
\centerline{\includegraphics[scale=0.35,trim=250 10 10 10]{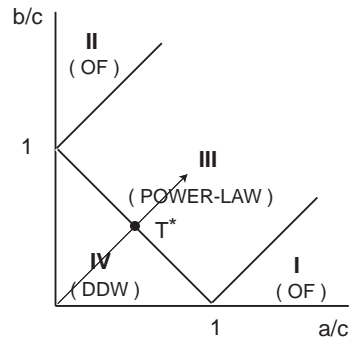}}
\caption{The phase diagram. The regions I and II are orbital ferromagnets (OF). The region III is the
power-law phase and the region IV is the orbital antiferromagnet with DDW order. The arrow marks a path (tetragonal
symmetry assumed) from the low temperature to the high temperature phase with the pseudogap transition at
$T_c=T^*$.}
\label{fig:phase}
\end{figure}

The phase diagram
is shown in Fig.~\ref{fig:phase}. The regions I and II have orbital ferromagnetic order due to macroscopic currents and
correspond to  $a> b+c$ and $b > a+c$, respectively. The region III is disordered, corresponding to $a, b,
c<\frac{1}{2}(a+b+c)$, but the current-current correlation function
exhibits a power-law. It includes the infinite temperature case $a=b=c=1$. This entire high temperature region
is on the the critical line of the 8-vertex model\cite{Lieb,Baxter}. The region IV is the DDW phase, which
is an orbital antiferromagnet  and corresponds to $c>a+b$. The phase boundary between region III and IV is given
by $\frac{b}{c}=-\frac{a}{c}+1$, or $c=a+b$, which implicitly defines the  transition temperature $T_c$.

Consider a given set values of $a$, $b$, and $c$. As the temperature
increases from 0 to $\infty$, this point follows a path always ending at $(1,1)$. This path may or not
cross a phase boundary. The path followed in
the phase diagram  when
$a=b$ is shown as an arrow. On this path, at $T_c=T^*$, there is a phase transition from the ordered
DDW phase to the disordered high temperature phase with a power-law correlation in the current-current
correlation function. The temperature $T^*$ is given by
\begin{equation}
T^* = \frac{\varepsilon_1-\varepsilon_5}{\ln 2},
\end{equation} 
determining,  phenomenologically, the energy difference $\Delta\varepsilon =\varepsilon_1-\varepsilon_5$.

In region IV, the correlation length, $\xi$, is\cite{Baxter,McCoy} 
\begin{equation}
\xi=-\ln \left[2x^{1/2}\prod_{m=1}^{\infty}\left(\frac{1+x^{4m}}{1+x^{4m-2}}\right)\right],
\end{equation}
where $x=e^{-\lambda}$ and $-\cosh \lambda = (a^2+b^2-c^2)/2ab$. As $T\to T_c^-$, $\lambda\propto (T_c-T)^{1/2}$, and it
can be seen by applying Poisson summation formula that
\begin{equation}
\xi^{-1}=8e^{-\frac{\pi^2}{2\lambda}}, \; \lambda \to 0.
\end{equation}

\subsection{Free energy and specific heat}
The exact free energy of the 6-vertex model was obtained by Lieb\cite{Lieb1}, and it is quite remarkable. If we denote
the free energy density in region III by $f_{\text{III}}(T>T_c)$, then its analytic continuation to $T<T_c$ is complex,
equal to $f_{\text{IV}}(T<T_c)+if_{\text{sing}}$.  The singular part of the free
energy has only an essential singularity at
$T=T_c$, which is given by
\begin{equation}
f_{\text{sing}}\propto \xi^{-2}.
\end{equation}
In fact,  all temperature derivatives of the free energy exist and are identical on both sides of the transition. An
asymptotic expansion at $T_c$, which is the same above and below the transition, is given in Ref.~\cite{Lieb}. 

The essential
singularity  implies that there are no observable specific heat anomalies, as any
derivative of the free energy vanishes at this infinite order transition. The situation is exactly the same as that of
the Kosterlitz-Thouless phase transition of the two-dimensional XY-model. The transition of the 6-vertex model is
inverted, however, as the low temperature phase has a finite correlation length, while the the high temperature phase
is the power law phase with continuously varying critical exponent. The analytic part of the free energy  results in 
a broad heat capacity peak above the transition\cite{Chakravarty5} for the XY-model, but below the transition in the
6-vertex model!

\begin{figure}[htb]
\centerline{\includegraphics[scale=0.35]{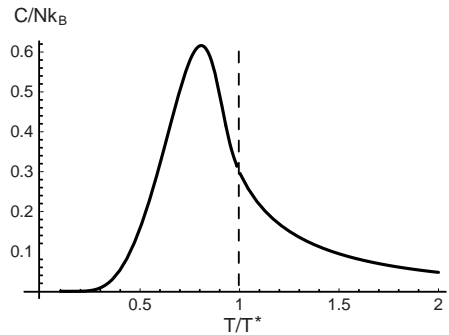}}
\caption{The specific heat of the 6-vertex model as a function of $T/T^*$ for $a=b$. Note that there are no observable
anomalies at the transition at $T^*$, and the peak of the specific, which is entirely due to the analytic part of the
free energy, is significantly below the transition.}
\label{fig:heat}
\end{figure}

The specific heat of the 6-vertex model was also discussed in Ref.~\cite{Lieb}, but it is
illuminating to plot it here.  In Fig.~\ref{fig:heat}, we show $C/Nk_B$, where $N$ is the number of vertices,
as a function of
$T/T^*$, where we have set $a=b$. There is no observable anomaly at $T^*$ despite the thermodynamic transition at this
temperature to the  ordered  DDW phase. 
Nonetheless, since  this is an order-disorder transition, the high
temperature entropy must disappear as $T\to 0$. At $T=\infty$, $a=b=c=1$, and the entropy is that of square ice, which
was calculated by Lieb\cite{Lieb2} to be $S=\frac{3k_B}{2}\ln \frac{4}{3}$. The system looses this entropy  by a
specific heat hump clustered around a definite temperature, which is significantly below $T^*$. The specific heat per
vertex is also a universal function of $T/T^*$. That it vanishes at high tempeartures --- true for many
statistical mechanical models such as the Ising model--- simply reflects the bounded energy spectrum on the high energy
side and must not be taken seriously. The high energy spectrum is not modeled correctly by such low energy effective
models. As a result of the boundedness of the energy spectrum, the entire passage from  complete order to complete
disorder takes place over a limited range of temperature. However, we  should probably take seriously the fact that the
DDW transition results in a specific heat of order $\sim 0.5 k_B$ per vertex considerably below the DDW transition and
may be mistaken as the pretransitional effect as the system enters the superconducting state from the pseudogap phase.
The specific heat peaks at the superconducting transition in the underdoped regime indeed appear to be anomalous and
very broad.  The low temperature behavior of the specific heat is given by 
\begin{equation}
\frac{C}{Nk_B}=2 \left(\frac{ T^*\ln 2}{ T}\right)^2 e^{-\frac{ T^*}{ T}\ln 2}, \; T\to 0.
\end{equation}

\subsection{Floating power-law phase}
 How does the coupling between the two-dimensional CuO-planes change the behavior of the system? Since the low
temperature phase is ordered with a finite correlation length, small coupling between the planes will in general have
small effect, except close to the critical region where the correlation length diverges and a crossover to 3D should
take place. This can be checked explicitly from a mean field analysis. The coupling between the planes due
to the orbital currents is indeed very small, as the magnetic field at the center of a plaquette generated by the
circulating currents can be estimated to be of order 10 G from the magnitude of the pseudogap\cite{Chakravarty2}. There
is some experimental support to this fact\cite{Mook}, as the correlation length in the perpendicular direction does not
extend much beyond a unit cell.

In contrast, the effect of  three-dimensional (3D) coupling in the high-temperature power-law phase is much more
interesting. In principle, the 2D power law phase can be destroyed by arbitrarily weak coupling between the planes.
However, because of the miniscule coupling between the planes due to an estimated 10 G magnetic field of a
plaquette\cite{Chakravarty2}, the crossover to 3D order is likely to take place at very long length scales. Thus, this is
effectively a floating phase in which the system behaves as a stack of decoupled 2D power-law phases. The argument, in principle,
is  the same as that  of Josephson coupled 2D XY-systems in a direction perpendicular to the
planes\cite{Lubensky}, although the details are different. 

Consider the current-current coupling of a given layer, $n$, with a neighboring layer $(n+1)$, in the continuum limit, which is
\begin{equation}
H_c[n] = g \int \frac{d^2r}{s^2} {\mathbf j}_n(r)\cdot  {\mathbf j}_{n+1}(r),
\end{equation}
where $s$ is a short distance cutoff of the order of the lattice spacing. It is sufficient to consider the coupling
between the nearest neighbor planes for our purposes.  We wish to determine the relevancy of the coupling $g$ under the
renormalization group transformation. Let 
\begin{equation}
\langle {\mathbf j}_n(r)\cdot  {\mathbf j}_{n+1}(r) {\mathbf j}_n(0)\cdot  {\mathbf j}_{n+1}(0)\rangle
\sim\frac{1}{r^{2x}},
\label{eq:correlation}
\end{equation}
where the average is taken with respect to the fixed point Hamiltonians of the uncoupled planes. Since every
point in the power-law phase  corresponds to a fixed point with a continuously varying critical exponent, we merely
need to determine the critical exponent $x$ of the current-current correlation function. 

The exponent of the arrow-arrow correlation function of the 6-vertex model (or the current-current correlation function in the
present problem) in the power-law phase can be obtained from the exponent of the 8-vertex model along the critical line, which in
turn can be obtained from the
$S_z$-$S_z$ correlation function of the one-dimensional (1D) quantum XXZ Hamiltonian. The correspondence can also be seen  by
considering  Trotter-Suzuki  decomposition of the 1D XXZ Hamiltonian, in which the arrow correlations along
a row of the 6-vertex model (or along a column) correspond to correlations  along the diagonal of the space-imaginary time
lattice\cite{Barma}. This subtlety is of no consequence, as we are interested in the exponent along a critical line at which the
correlation length is infinite.

The $S_z$-$S_z$ correlation has a staggered part and a uniform part. The exponent for the uniform part is 2, independent of
temperature, while the exponent of the staggered part is given by $\frac{1}{\theta}$, where\cite{McCoy,Luther}
\begin{equation}
\theta= 1- \frac{\mu}{\pi}.
\end{equation}
Assuming tetragonal symmetry ($a=b$),
\begin{equation}
\cos \mu = \frac{1}{2}e^{(2\ln 2)\frac{T^*}{T}} - 1 .
\end{equation}
Therefore  the exponent $\frac{1}{\theta}$ varies monotonically from 1 at $T=T^*$ to $3$ at $T=\infty$. The exponent of the
uniform part follows from a local conservation law combined with the conformal invariance in two dimensions\cite{Ginsparg}. Clearly,
the slowest decay will determine the 3D crossover. At
$T^*$, it is determined by the staggered part since its exponent is less than the exponent  of the the uniform part. As the
temperature increases, the exponent of the staggered part increase and crosses the uniform exponent. We therefore define the
exponent
$x$  to be  either the exponent of the uniform part, or that of the staggered part of the current-current correlation function,
whichever is smaller. Under a renormalization group transformation, for which
$s\to\lambda s$, where
$\lambda >1 $ is a scale factor, ${\mathbf j}_n(r)\cdot  {\mathbf j}_{n+1}(r)\to
\lambda^x {\mathbf j}_n(r)\cdot  {\mathbf j}_{n+1}(r)$ from Eq.~\ref{eq:correlation}. Thus, for the free energy to
remain invariant, it follows that $g\to \lambda^{2-x} g$; note  that $d=2$. Therefore,  at $T=T^*$, where
$x=1$, the 3D coupling, $g$, is strongly  relevant. As the temperature $T\to \infty$, the unfiorm exponent ($x=2$) takes over and
$g$ becomes marginal.

\section{Quantum vertex model}

It is interesting to ask what a quantum generalization of this model could be.  At the very least, we  must allow the bond currents
to flip as a result of quantum fluctuations. The most local move  is to  flip the currents around an elementary plaquette of the
square lattice. But it is easy to see that such a process can   create sources and sinks, if we do not impose any constraints.
From a path integral description, the configuration at an intermediate (imaginary) time slice  may
contain a source or a sink even though the states at the initial and the final time slices at 0 and
$\beta$ do not. 

I start by considering the 8-vertex model written in terms of Ising spin
variables and then incorporate quantum mechanics by introducing  a transverse
field on a dual lattice site. The advantage is that the
description is in terms of the familiar Ising basis\cite{Note}.  One must then  explicitly introduce projection
operators to eliminate the unwanted  sources and sinks. 

\begin{figure}[htb]
\centerline{\includegraphics[scale=0.35]{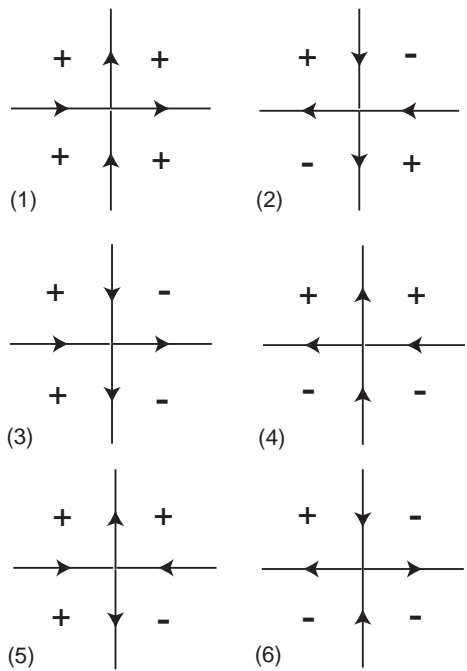}}
\caption{The mapping of the first 6 vertices to equivalent spins on the dual lattice.}
\label{fig:spin-vertex1}
\end{figure}

\begin{figure}[htb]
\centerline{\includegraphics[scale=0.35]{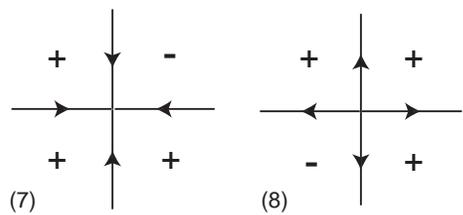}}
\caption{The mapping of the vertices involving sinks and sources to spins on the dual lattice.}
\label{fig:spin-vertex2}
\end{figure}

The 8-vertex model (allowing only an even number of arrows into and out of each site) can be written as 
two coupled Ising models on  interpenetrating square lattices with an additional four spin coupling, as is
well-known\cite{Kadanoff,Wu}. The spins are situated on the dual lattice consisting of the sites  at the center of
the plaquettes of the original square lattice of bond currents. An arrow to the right (or upward) represents 
adjacent  parallel spins; similarly, an arrow to the left (or downward) corresponds to adjacent antiparallel spins.
The mapping is shown explicitly in Figs.~\ref{fig:spin-vertex1} and
\ref{fig:spin-vertex2}. The additional vertices shown in Fig.~\ref{fig:spin-vertex2} are assigned energies
$\varepsilon_7=\varepsilon_8$, and the number of such vertices must satisfy $n_7=n_8=n_d$. We shall denote these
vertices as of type $d$. However, ultimately, we must impose an additional constraint on the Hamiltonian in which the
vertex configurations shown in Fig.~\ref{fig:spin-vertex2} are removed  from the Hilbert space of states.

\begin{figure}[htb]
\centerline{\includegraphics[scale=0.35]{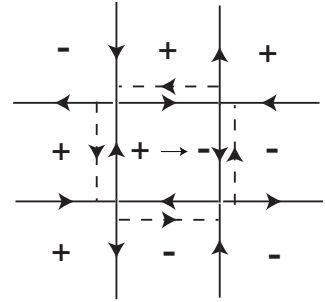}}
\caption{The flip of the central spin from $+\to -$ flips the surrounding bound currents around the plaquette. The
flipped currents are denoted by dashed arrows.}
\label{fig:spin-flip}
\end{figure}

Quantum fluctuations are incorporated by a transverse
field, $h$. The resulting model, the quantum 6-vertex model,  is then defined by the
following partition function:
\begin{equation}
Z_Q=\text{Tr}\;  \; e^{-\beta P H P} ,
\end{equation}
where the Hamiltonian $H$ is given by
\begin{eqnarray}
H= -h\sum_{j,k}\sigma^x_{j,k}&-&
\sum_{j,k}\left(J\sigma^z_{j,k}\sigma^z_{j+1,k+1}+J'\sigma^z_{j+1,k}\sigma^z_{j,k+1}\right)\nonumber \\
&-&J''\sum_{j,k}
\sigma^z_{j,k}\sigma^z_{j+1,k+1}\sigma^z_{j+1,k}\sigma^z_{j,k+1}.
\end{eqnarray}
The operators $\sigma^x$ and $\sigma^z$, are the standard Pauli matrices, and
$P^2=P$ is the projection operator that projects out the sources and sinks. It is easy to check that a single spin flip
at a dual lattice site flips the  bond currents on the surrounding plaquette, which causes transitions between the
vertices on the four corners as shown in Fig.~\ref{fig:spin-flip}. The previously defined vertex weights $a$, $b$, and
$c$ are given by 
\begin{eqnarray}
a&=& e^{(J+J'+J'')/T},\\
b&=& e^{(-J-J'+J'')/T}, \\
c&=& e^{(-J+J'-J'')/T}.
\end{eqnarray}
The new vertex weight $d$ is
\begin{equation}
d=e^{(J-J'-J'')/T}.
\end{equation}

The classical statistical mechanics that controls the finite temperature phase transition is unchanged
because there are no quantum mechanical spin flips. Since the trace is a sum over   states from which sources and sinks are
projected out, the model is clearly the classical  6-vertex model---the number of vertices of type
$d$, $n_d =0$. We can  arrive at the same conclusion by considering an imaginary time path integral from a Trotter decomposition
along the imaginary time direction, $\tau$, between $0$ and $\beta$, with periodic boundary condition. The constraint can be
enforced by including a delta functional
$\delta(\mathbf{\nabla}\cdot\mathbf{j(x,\tau))}$ at each space-time point. In the classical limit, we are left with a single time
slice, because quantum spin flips due to the transverse field are punished. The partition function is now a sum over all spin
configurations subject to the condition that there are no sources or sinks. Of course, at any finite temperature, the effective
parameters of the model will be renormalized by quantum fluctuations at short distances, but that can not change the universality
class of the finite temperature phase transition.

At $T=0$, the model is  somewhat complex  because of the projection operators.  At 
$h=0$, the ground state is the ordered DDW state with no quantum fluctuations,
if the vertices
$c$ have the lowest possible energy. As the transverse field
$h$ increases from zero,  quantum fluctuations increase and nucleates PDW vertices. It is
not clear what the ultimate fate of the system is. There are four distinct possibilities: (1) a quantum phase transition
to a gapped quantum disordered state, which I find unlikely because of the constraints. If this is the case, the constraints have to
be somehow irrelevant; (2) a transition to a PDW state, which I also find unlikely because there are no flippable
plaquettes, and hence the state is disfavored from kinetic energy considerations; (3) a transition to a  power-law phase
since a local deformation may not be able to heal sufficiently fast. (4) There is also a remote possibility that the system does not
loose its DDW order. In any case,  the model is sufficiently complex that at present I cannot provide a quantitative
analysis. It is probably best studied by quantum Monte Carlo methods involving loop algorithms\cite{Evertz}. 

\begin{figure}[htb]
\centerline{\includegraphics[scale=0.35]{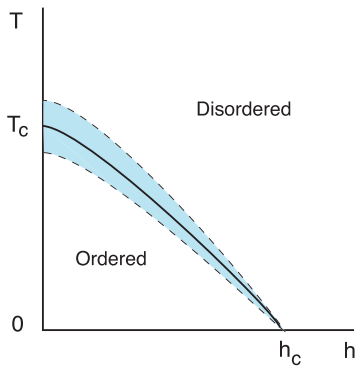}}
\caption{Conventional phase diagram in which a finite temperature phase transition exists --- the solid curve. The
shaded region is the domain of classical critical fluctuations. The influence of the quantum critical point at $h_c$, at
$T=0$, has to lie outside this domain.}
\label{fig:QCP1}
\end{figure}

I shall assume that it is the third possibility discussed above that is realized. If this is correct, such a quantum critical point
is unusual from the conventional perspective\cite{Hertz}. To illustrate this point, consider a phase diagram in which a system
undergoes a continuous phase transition as a function of tuning parameters, temperature,
$T$, and a coupling $h$, such as the transverse field. This is shown in Fig.~\ref{fig:QCP1}, where the shaded region is the domain
of classical critical fluctuations extending all the way to zero energy. Any influence of the quantum critical point at $h_c$ has to
be outside this domain. So, we may argue that, on some scale, the disordered phase at a finite temperature could be influenced by
the quantum critical point.

\begin{figure}[htb]
\centerline{\includegraphics[scale=0.35]{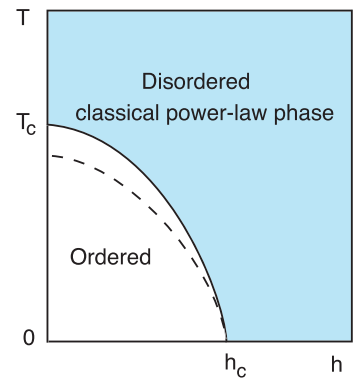}}
\caption{The phase diagram in which the entire finite temperature disordered phase is a classical power-law phase ---
the shaded region. The dashed curve is the crossover scale for the classical critical fluctuations within the ordered
phase. The quantum critical regime is entirely eliminated, except precisely at $T=0$. The ordered regime is 
under the influence of the stable broken symmetry fixed point.}
\label{fig:QCP2}
\end{figure}

The phase diagram shown in Fig.~\ref{fig:QCP2}, pertinent to our present problem,  is strikingly different. Here, the
existence of a classical power-law phase of the 6-vertex model has entirely eliminated the quantum 
fluctuations, which are, strictly speaking, cutoff by the thermal length, while the classical power-law correlations have no such
long-distance  cutoff. Within the ordered phase, there is a classical critical region, but otherwise the ordered phase is under the
influence of the broken symmetry fixed point. Of course, we are assuming that the magnitude of the order parameter can be assumed
to be fixed. If not, there will be a further transition at a higher temperature where its magnitude collapses. The region above
this transition can now be influenced by the quantum critical point. Nonetheless, there may be a sizable region of the phase
diagram in which the classical power-law phase dominates.

\section{Relation to the quantum dimer model}

As remarked in the Introduction, there are some similarities with the RVB ideas, particularly with the recent
developments involving quantum dimer models\cite{Sondhi,Sondhi2}, but the distinctions are equally important. At high
temperatures, the quantum dimer model on a square lattice maps on to the classical dimer model with power-law
correlations\cite{Fisher}, which  is a $c=1$ (conformal charge) theory, equivalent to two massless Majorana fermions. 
It is also the fully frustrated Ising model at $T=0$.

This critical phase arises
from the Rokhsar-Kivelson quantum critical point\cite{Kivelson,Henley}. Instead, in the present case, the quantum vertex
model maps on to the classical 6-vertex model, which undergoes a phase transition with an essential singularity from an
ordered  region IV to the power-law  region III. But this power-law phase  does not stem from a
quantum critical point, instead the entire  region is situated on the critical line of the classical 8-vertex model. 

Recent thrust
in quantum dimer models has focused on the existence of a spin liquid phase, which is made plausible by the fact that
the classical dimer model on a triangular lattice is non-critical with exponentially decaying dimer
correlations\cite{Sondhi}. Thus, the ground state of the triangular case has no gapless collective excitations, only
deconfined, gapped spinons for a finite range of parameters. In contrast, we have focused on the existence of a
power-law phase above the DDW phase and the non-existence of the specific heat anomaly at the transition. We have also
argued that it is also possible to have a quantum phase transition in the ground state, which is induced by a
sufficiently large transverse field mimicking the actual pseudogap transition in the groundstate. There are also 
interesting resemblances of our work  with the quantum  and sliding ice pictures of Ref.~\cite{Sondhi2}.

\section{Vertex models on a triangular lattice}

Vertex models may be useful in other contexts. For example, consider the
ideas that were put forward recently regarding the metal-insulator transition in two-dimensional electron
systems in which the transition  is identified as a quantum phase transition in a disordered Wigner
crystal state\cite{Chakravarty3,Chakravarty4}. The chain of reasoning is as follows: extensive calculations of multiparticle
exchange Hamiltonians in  pure\cite{Chakravarty3,Chakravarty4,Ceperley} and disordered Wigner crystals\cite{Chakravarty4} suggest
that there exist ground   states with well-defined bond currents. These bond currents can be represented by vertex models on a
triangular lattice. It is notable that recent studies have revealed that a metal-insulator transition could occur in such models. A
protypical example is provided by  a two dimensional tight-binding  Hamiltonian  in a random magnetic field and on-site
disorder\cite{Sheng,Hoang}. It is in this context that it is useful to consider a vertex model of bond currents
on a triangular lattice and study its quantum dynamics. 
Although even in the classical limit, such  models are  unsolved in general\cite{Baxter,Baxter2,Kelland}, some exact
results, known for certain special cases,  are sufficiently encouraging and similar to the 6-vertex model to pursue further.

\begin{figure}[htb]
\centerline{\includegraphics[scale=0.35]{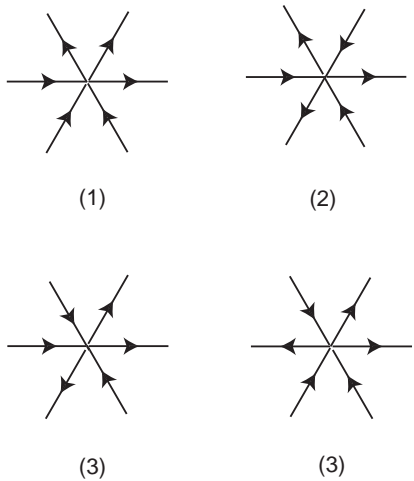}}
\caption{Examples of three types of allowed vertices.}
\label{fig:triangle}
\end{figure}

If we restrict
ourselves to 3 incoming currents and 3 outgoing currents at a site of a triangular lattice, this generates a $6!/3!3! = 20$ vertex
model. If the vertices which differ only by rotation and reflection are treated alike, they may be classified as follows: (i) 6
vertices in which the incoming arrows are adjacent, (ii) 2 vertices in which the incoming and the outgoing arrows
alternate round the vertex, (iii) 12 vertices containing two incoming arrows directly opposite each other.  The examples
are shown in Fig.~\ref{fig:triangle}.

Unlike the 8-vertex, or the 6-vertex, model on a square lattice, the solution to the 20-vertex model is not known in
general. It is only solved for certain values of the vertex configuration probabilities. As
with the 6-vertex model, this model undergoes a phase transition to a power-law ordered phase, with an essential
singularity in the free energy, from an ordered state with finite correlation length in which the arrows alternate along
a row. Since the exact solution is not known in general, it would be extremely useful explore this model
numerically\cite{Evertz}.

The quantum dynamics of this model can once again be implemented by going over to a Ising spin model on the dual
lattice with a transverse field at a site. It would be interesting to explore this model,  which could provide us with a
novel perspective on the old problem of a Wigner crystal, which, in the conventional picture, is an ordered
state  with at most small zero-point oscillations of an electron about a lattice site; see, however,
Refs.~\cite{Chakravarty3,Chakravarty4}. 

\section{Conclusions}

I have argued that the natural generalization of the Hartree-Fock model of the DDW are certain vertex models
involving bond currents. The finite temperature transition from the ordered DDW phase to the disordered phase belongs 
to the universality class of the 6-vertex model and  is striking because of the absence of any observable specific heat
anomalies. I have also shown that the disordered phase is a power-law phase of the bond current correlations. These
power-law correlations do not arise from any quantum critical point, but this entire phase is poised on the critical
line of another hidden model, the 8-vertex model. The origin of the power-law correlations  is the constraints that the
configurations of the 6-vertex model must satisfy. Thus, a local deformation cannot heal within a finite correlation
length. In this respect, there is a strong similarity with the RVB dimer models on a square lattice, where the
constraints in the dimer model are once again responsible for the high temperature power-law correlations. 

It is important to note that the mere knowledge of the nature of the
order parameter does not necessarily determine the universality class of the phase transition. For this, it is
necessary to know also the nature of the possible excitations that can be thermally populated. Thus, labelling the DDW
transition as an Ising transition is an oversimplified view.

I was also able to construct a quantum generalization involving a 8-vertex model in  a transverse field, but with projection
operators that ultimately project out the sources and the sinks of the 8-vertex model. The model is sufficiently complex to present
any  in-depth discussion at this time, but it is not difficult to motivate a quantum phase transition as a function of the
transverse field. Nonetheless, it is remarkable that the effect of the quantum critical point is completely swamped by the
classical power-law correlations at any finite temperature. This is a strict departure from the conventional theory of quantum
critical point\cite{Hertz}.

I have also speculated on the possible application of a similar vertex-model to  the Wigner crystal  and its
implications to the metal-insulator transition in the two-dimensional electron systems. Similarly, a zoo of new sets of
density wave theories\cite{Nayak} can be usefully formulated in terms of vertex models opening up a whole class of
questions in the theory of quantum statistical mechanics. Some of these would correspond to modulations of the bond
kinetic energy\cite{Sachdev,Nayak}. In fact, it is quite possible that
certain regions of the cuprate phase diagram are  described by these models\cite{Sachdev2}, but the aim here was to
consider only those models that are natural generalizations of the proposed DDW state.

The present analysis is incomplete in one important respect. I have been unable to make firm statements regarding the
quasiparticle excitations and doping in the present model, although it is clear that  power-law bond current
correlations in the disordered phase probably leads to non-fermi liquid behavior of the electronic excitations. Since, we
have considered models of the spin-singlet variety, there cannot be any spin-charge separation by definition. The
non-fermi liquid behavior must then be akin to that of the spinless Luttinger model in one dimension. One should note
that a rigorous description of  electronic excitations are also unavailable for the RVB quantum dimer model.

In the future, it may also be interesting to allow for sources and sinks, as they may represent an effective means of incorporating
decay of collective excitations involving the order parameter into low lying nodal quasiparticles in the actual physical
system\cite{Kivelson2}. From this perspective, the correct quantum model is actually a 8-vertex model in a transverse field, as
discussed above, but without any complications of the projection operators.

\acknowledgments
This work
was supported by a grant from the National Science Foundation: NSF-DMR-9971138. I thank E. Demler, J. Fjaerestad, C. L. Henley, J.
P. Hu, H. -Y. Kee, J. B. Marston, C. Nayak, C. Panagopoulos, S. Sachdev, S. L. Sondhi, and S. Tewari for
many interesting comments. I am especially grateful to S. Kivelson for his insightful comments on all aspects of this work.

\end{document}